\documentclass[aps,prd,reprint]{revtex4-2}

\usepackage{bm}
\usepackage{amssymb}
\usepackage{epsfig}
\usepackage{slashed}
\usepackage{color}
\usepackage{mathtools}
\usepackage{cases}
\usepackage{booktabs}
\usepackage{amsmath}
\usepackage{url}
\usepackage{graphicx}
\usepackage[
colorlinks=true,
filecolor=black,
anchorcolor=blue,
linkcolor=blue,
citecolor=cyan, 
urlcolor=cyan,
linktocpage=true,
plainpages=false,
breaklinks=true,
pdfstartview=FitH
]{hyperref}

\definecolor{deepblue}{rgb}{0,0.3,0.8}

\def\GW{\text{GW}}
\def\dd{\mathrm{d}}

\def\zz{\zeta}
\def\Mp{M_P}
\def\fh{f_{\alpha}}
\def\A{\mathcal{A}}
\def\kk{\vec{k}}

\newcommand{\bea}{\begin{eqnarray}}
\newcommand{\eea}{\end{eqnarray}}
\newcommand{\beq}{\begin{eqnarray}}
\newcommand{\eeq}{\end{eqnarray}}

\begin{document}

\title{Chiral Gravitational Wave Background from Audible Axion via Nieh-Yan Term}

\author{Baoyu Xu$^{1,3}$}
\author{Keyi Ding$^{4}$}
\author{Hong Su$^{2,6,7}$}
\author{Ju Chen$^{5,6}$}
\author{Yun-Long Zhang$^{1,2}$}
\email[Corresponding author:~]{ zhangyunlong@nao.cas.cn}

\affiliation{$^1$National Astronomical Observatories, Chinese Academy of Sciences,
Beijing 100101, China}
\affiliation{$^2$School of Fundamental Physics and Mathematical Sciences, Hangzhou Institute for Advanced Study, University of Chinese Academy of Sciences, Hangzhou 310024, China}

\affiliation{$^3$School of Astronomy and Space Science, University of Chinese Academy of Sciences, Beijing 100049, China
}
\affiliation{$^4$Van Swinderen Institute for Particle Physics and Gravity, University of Groningen, Nijenborgh 4, 9747 AG Groningen, The Netherlands
}

\affiliation{$^5$International Center for Theoretical Physics Asia-Pacific (ICTP-AP), University of Chinese Academy of Sciences, Beijing 100190, China
}

\affiliation{$^6$Taiji Laboratory for Gravitational Wave Universe (Beijing/Hangzhou), University of Chinese Academy of Sciences, Beijing 100049, China
}

\affiliation{$^7$CAS Key Laboratory of Theoretical Physics, Institute of Theoretical Physics,
Chinese Academy of Sciences, Beijing 100190, China
}

\begin{abstract}
Axions and axion-like particles can be probed through gravitational waves indirectly, often referred to as ``audible axions''. The usual concept of audible axion relies on the coupling between the axions and the gauge fields. Here we consider an axion-like mechanism with coupling to the Nieh-Yan term. This interaction leads to the direct and efficient production of gravitational waves during the radiation-dominated era, originating from the tachyonic instability of the gravitational perturbations with the Nieh-Yan term. We calculate the energy spectral density of the chiral gravitational
wave background and the comoving energy density of axion-like fields. Based on the numerical results, we explore the parameter space of axion masses and decay constants for detectable gravitational wave signals, either in pulsar timing arrays or space-based gravitational wave detections.
\end{abstract}

\maketitle
\tableofcontents

\section{Introduction}
\label{introduction}
In modern physics, understanding the nature of dark matter is crucial. One of the candidates for dark matter is the axion~\cite{Preskill:1982cy,Abbott:1982af,Dine:1982ah,Bertone:2004pz, Bertone:2016nfn, Ferreira:2020fam, Hui:2021tkt, Antypas:2022asj,freese1990natural}, in general, it also includes axion-like particles (ALPs)~\cite{ Ferreira:2020fam, Hui:2021tkt, Antypas:2022asj, Moroi:2020has, Sun:2021yra,An:2020jmf,Adams:2022pbo,Choi:2020rgn}.
The axion is originally introduced to solve the strong CP problem~\cite{Peccei:1977hh,Wilczek:1977pj}, and it also appeared extensively in other studies such as string theory~\cite{Svrcek:2006yi}. Following the first detection of gravitational waves by LIGO~\cite{LIGOScientific:2016aoc}, the detection of new particles in cosmology, such as axions or ALPs, by gravitational waves has become possible~\cite{Brito:2017wnc,Machado:2018nqk,Madge:2021abk,Gerlach:2025fkr,Jiao:2025xnz}. In recent years, many studies focusing on parity-violating gravity,  with Chern-Simons modified gravity being one of the most notable examples~\cite{Jackiw:2003pm,Alexander:2009tp,Bombacigno:2022naf,Kawai:2017kqt,Peng:2022ttg,Nojiri:2019nar,Nojiri:2020pqr,Chu:2020iil,Sun:2020gem,Li:2023vuu,Ding:2024}. In this work, we utilize another healthy parity-violating model known as Nieh-Yan modified teleparallel gravity (or Nieh-Yan gravity)~\cite{Li:2020xjt,Li:2021wij,Cai:2021uup,Wu:2021ndf,Li:2023fto,Li:2022vtn,Rao:2023doc,Zhang:2024vfw,Li:2022mti}. 

{The Nieh-Yan gravity is a theory that modifies the teleparallel equivalent of general relativity (TEGR) with the dyanmical Nieh-Yan term~\cite{nieh1982identity,Nieh:2008btw}, which is coupled with the axion or axion-like field.} We explore the different parameter regions built with axion masses and decay constants, and we numerically calculate the chiral gravitational wave background~\cite{Caprini:2018mtu,regimbau2011astrophysical,Thrane:2013oya,Cai:2016ihp,Bartolo:2018elp,Odintsov:2022hxu}  produced by the audible axion via Nieh-Yan term.  We find that the gravitational wave spectral energy density varies with different helicities. Considering the evolution of gravitational waves, we identify detectable signals in pulsar timing arrays (PTA)~\cite{NANOGrav:2023icp,Reardon:2023gzh,EPTA:2023fyk,Xu:2023wog,Belgacem:2020nda} or future space-based gravitational wave detectors such as LISA~\cite{Barausse:2020rsu}, Taiji~\cite{Ruan:2018tsw}, and ASTROD-GW~\cite{Ni:2012eh}. 

Furthermore, by calculating the axion comoving energy density within the axion dynamics which is affected by gravitational backreaction, we find that backreaction of the Nieh-Yan term can reduce the axion comoving energy density by several orders of magnitude. Thus, in some parameter spaces, the relic density of axion dark matter can be comparable to the observational results~\cite{Venegas:2021wwm,Arias:2022qjt}, and the chiral gravitational wave radiation energy density is also within a reasonable range~\cite{Planck:2018vyg}.

The organization of this paper is as follows. In section~\ref{sec2}, we briefly review the TEGR, introduce the Nieh-Yan modified teleparallel gravity, and discuss how the axion dynamics are affected by geometric backreaction. In section~\ref{GWP}, 
we study the production mechanism of chiral gravitational waves via Nieh-Yan term, by numerically calculating the energy spectral density of chiral gravitational waves. The process of numerical calculation is described in detail. In section~\ref{detection}, we explore the parameter region of axion mass and decay constants that could be detected by PTA and the space-based gravitational wave detectors. The time evolution of the comoving energy densities of the axion and the gravitational wave are calculated. In the last section~\ref{conclusion}, we conclude with the discussions.  Throughout this paper, we choose the signature difference as $(-,+,+,+)$, and set $c=\hbar=1$.


\section{Audible axion with Nieh-Yan term}
\label{sec2}
In this section, we overview the TEGR and Nieh-Yan modified teleparallel gravity. The TEGR was initially proposed by Einstein as an attempt to unify gravitational and electromagnetic interactions~\cite{Hohmann:2022mlc,Maluf:2013gaa,Bahamonde:2021gfp}.
\subsection{Teleparallel equivalent of general relativity}
{The TEGR is based on the theory of tetrad-spin connection pairs.} In this framework, the fundamental variables of spacetime are represented by the tetrad $e^{A}_{~\mu}$ and the spin connection $\omega^{A}_{~B\mu}$. This allows the metric to be written as
\begin{equation}
    g_{\mu\nu} = \eta_{AB} e^{A}_{~\mu}e^{B}_{~\nu},
\end{equation}
{and the metric determinant can be replaced by the tetrad determinant through $e=\sqrt{-g}$.}

TEGR is a dynamically equivalent theory to general relativity,  sharing the same field equations and classical predictions as general relativity, and can address similar questions~\cite{Combi:2017crv}, {such as Mercury's perihelion precession}. In TEGR, the curvature is replaced by torsion, resulting in a zero Riemann tensor:
\begin{align}\label{Riemann}
{R}^\sigma_{~\rho\mu\nu} = &\partial_\mu {}\Gamma^\sigma_{~\rho\nu} - \partial_\nu {}\Gamma^\sigma_{~\rho\mu} +  {}\Gamma^\sigma_{~\alpha\mu} {}\Gamma^\alpha_{~\rho\nu} - {}\Gamma^\sigma_{~\alpha\nu} {}\Gamma^\alpha_{~\rho\mu} \nonumber \\
=& {e^{~\sigma}_{A}e^{B}_{~\rho}[\partial_{\mu}\omega^{A}_{~B\nu}-\partial_{\nu}\omega^{A}_{~B\mu}+\omega^{A}_{~C\mu}\omega^{C}_{~B\nu}-\omega^{A}_{~C\nu}\omega^{C}_{~B\mu}]}\nonumber \\
=&0.
\end{align}
The dynamics of the spacetime is governed by the torsion tensor, defined as
\begin{equation}
    \hat{T}^\sigma_{~\mu\nu} =  {}\Gamma^\sigma_{~\mu\nu} - {}\Gamma^\sigma_{~\nu\mu} =
    {e^{~\sigma}_{A}(\partial_{\mu}e^{A}_{~\nu}-\partial_{\nu}e^{A}_{~\mu})} \neq 0.
\end{equation}

\label{TEGR}
In this study, we focus on the TEGR action, which resembles the action of general relativity:
\begin{equation}\label{eq5}
    S_{\text{TEGR}}=\frac{\Mp^{2}}{2} \int \dd^{4} x \sqrt{-g} \mathring{R},
\end{equation}
where $\Mp \approx 2.4 \times 10^{18} \, \text{GeV}$ is the reduced Planck mass.
$\mathring{R}^{\rho}_{~\lambda\mu\nu}$ is related to the vanishing Riemann tensor \eqref{Riemann} through
\begin{equation}\label{RT1}
 {}{R}^{\rho}_{~\lambda\mu\nu} =\mathring{R}^{\rho}_{~\lambda\mu\nu} + P^{\rho}_{~\lambda\mu\nu} = 0.
\end{equation}
Here $P^{\rho}_{~\lambda\mu\nu}$ is the Riemann contortion tensor, which can be considered a canceling term that makes ${}{R}^{\rho}_{~\lambda\mu\nu}$ vanish. 
Similar to the Riemann tensor, $\mathring{R}^\rho_{~\lambda\mu\nu}$ can be expressed in the following form:
\begin{equation}\label{RT0}
    \mathring{R}^\rho_{~\lambda\mu\nu} = \partial_\mu \mathring\Gamma^\rho_{~\lambda\nu} - \partial_\nu \mathring\Gamma^\rho_{~\lambda\mu} +  \mathring\Gamma^\rho_{~\alpha\mu} \mathring\Gamma^\alpha_{~\lambda\nu} - \mathring\Gamma^\rho_{~\alpha\nu} \mathring\Gamma^\alpha_{~\lambda\mu}.
\end{equation}
{$\mathring{\Gamma}^\rho_{~\lambda\nu}=e_{A}^{~\rho}(\partial_{\lambda}e^{A}_{~\nu}+\omega^{A}_{~B\lambda}e^{B}_{~\nu})$} is the teleparallel connection, which is related to the Levi-Civita connection ${}{\Gamma}^{\rho}_{~\mu\nu}$ though
\begin{equation}\label{RT2}
 {}{\Gamma}^{\rho}_{~\mu\nu} =\mathring{\Gamma}^{\rho}_{~\mu\nu} + 
  K^{\rho}_{~\mu\nu}.
\end{equation}
Here $K^{\rho}_{~\mu\nu}$ is the teleparallel contortion tensor, it can be expressed as
\begin{equation}\label{RT3}
    K^{\mu}_{~\nu\rho} = \frac{1}{2} \left(\hat{T}_{\nu~\rho}^{~\mu}+\hat{T}_{\rho~\nu}^{~\mu}-\hat{T}_{~\nu\rho}^{\mu}\right).
\end{equation}

Using the relations \eqref{RT1}-\eqref{RT3} and following detailed calculations in~\cite{Hohmann:2022mlc, Bahamonde:2021gfp}, one finds
\begin{equation}\label{RT4}
    \mathring{R} = -\hat{T} + \hat{B},
\end{equation}
where $\hat{T} =  \frac{1}{4} \hat{T}^{\rho\sigma\mu}\hat{T}_{\rho\sigma\mu}+\frac{1}{2}\hat{T}^{\mu\sigma\rho}\hat{T}_{\rho\sigma\mu}-\hat{T}^{\rho}_{~\rho\sigma}\hat{T}^{~\mu\sigma}_{\mu}$ and $\hat{B} = -\frac{2}{e} \partial_{\rho} (e \hat{T}^{\mu\rho}_{~\mu})$. Since $\hat{B}$ is a surface term that does not affect the equations of motion, it can be neglected. The TEGR action in \eqref{eq5} is then simplified as
\begin{equation}
{S_{\text{TEGR}}} = -\frac{\Mp^{2}}{2}\int \dd^{4}x e \hat{T}.
\end{equation}
We can see that, unlike general relativity, the TEGR action depends only on the torsion scalar, which serves as the sole dynamical variable.
 
\subsection{Nieh-Yan modified teleparallel gravity}\label{model}
\label{s2}
Based on TEGR, a parity-violating gravity model without ghost instability has been studied in recent years~\cite{Li:2020xjt,Li:2021wij}, where the Nieh-Yan term was used to modify TEGR. The Nieh-Yan term is a topological density term and has been extensively studied in Riemann-Cartan theories~\cite{Nieh:1981kw}. It can be written as
 \begin{equation}
     S_{\text{N-Y}} = \int \dd^{4}x \sqrt{-g} \frac{\mathbb{C}}{4} (\hat{T}_{A\mu\nu} \widetilde{T}^{A\mu\nu}-\varepsilon^{\mu\nu\rho\sigma}{R}_{\mu\nu\rho\sigma}),
 \end{equation}
where $\hat{T}^{A}_{~\mu\nu}=e^{A}_{~\rho}\hat{T}^{\rho}_{~\mu\nu}$ and $\widetilde{T}^{A\mu\nu}=\frac{1}{2}\varepsilon^{\mu\nu\rho\sigma}\hat{T}^{A}_{~\rho\sigma}$,  $\mathbb{C}$ is the coupling coefficient. $\varepsilon^{\mu\nu\rho\sigma}=\epsilon^{\mu\nu\rho\sigma}/\sqrt{-g}$ being the Levi-Civita tensor, and $\epsilon^{\mu\nu\rho\sigma}$ is the antisymmetric symbol.

In the TEGR, the Riemann tensor is zero, which makes the existence of the Nieh-Yan term the only non-vanishing term. 
After coupling with the axion-like field $\phi$, the total action in our study is given by
\begin{equation}\label{total_equation}
\begin{split}
    S_{\text{total}} = \int \dd^{4}x \sqrt{-g}
\bigg(& -\frac{M^{2}_{\text{p}}}{2}\hat{T} -\frac{1}{2}\partial_{\mu} \phi \partial^{\mu}{\phi} - V(\phi) \\
& + \frac{\alpha \phi}{4{\fh}}  \hat{T}_{A\mu\nu} \widetilde{T}^{A\mu\nu} \bigg).
\end{split}
\end{equation}
We have introduced the axion dynamic,
and the coupling coefficient $\mathbb{C}=\alpha\phi/f_{\alpha}$,  
where $f_{\alpha}$ is the decay constant and $\alpha$ is the coefficient with dimension of $\Mp^2$. Analogous to the coupling between the axions and the gauge field~\cite{Machado:2018nqk,Madge:2021abk}, the total action shows that the presence of the Nieh-Yan term breaks the parity symmetry of the gravitational field.

In total action \eqref{total_equation}, we assume the cosine-like potential, which is derived from the lower-order approximations of QCD~\cite{GrillidiCortona:2015jxo},
\begin{equation}\label{potential}
    V(\phi) = m(T)^{2} \fh^{2}\left[1-\cos\left(\frac{\phi}{{\fh}}\right)\right].
\end{equation}
Here, $m(T)$ is the axion mass, which is often affected by temperature $T$~\cite{Agrawal:2017eqm,Hertzberg:2008wr}. 
In our study, we approximately neglect the effect of temperature on the mass of the axion, which means $m(T)=m$. 
Moreover, different theoretical frameworks predict various potential forms. For example, the full non-perturbative potential can be expressed as a Fourier expansion, and in some modified scenarios, non-periodic perturbative terms may appear~\cite{Choi:2020rgn}. However, when considering only the leading-order contribution, these potential forms can be well approximated by the one we have chosen. In particular, for the QCD axion mass  $m$ , the relation derived from chiral perturbation theory and lattice QCD simulations is as follows
~\cite{Gorghetto:2018ocs,Borsanyi:2016ksw}: 
\begin{equation}
    m = 5.69\times 10^{-3}\text{eV}
    \left(\frac{10^{9}\text{GeV}}{{\fh}}\right).
\end{equation} 
In this study, we use ${\fh} \sim 10^{17}\text{GeV}$ approximately, which allows us to neglect the abundance of thermal axions.

In the radiation dominated universe, the Hubble parameter $H(T)$ can be written as
\begin{equation}\label{H(T)}
    H(T)^{2} = \frac{\pi^{2}g_{\rho,*}(T)}{90\Mp^{2}}T^{4},
\end{equation}
{where $g_{\rho,*}$ is the number of relativistic degrees of freedom. Using this equation, we can define the oscillate begining temperature $T_{\text{osc}}$   by the approximately condition $H(T_{\text{osc}}) \simeq m$, we can get $T_{\text{osc}} \simeq \sqrt{m\Mp}$.}  
Moreover, this means that the axion begins oscillating around the QCD scale ~\cite{Agrawal:2017eqm}. The axion field is effectively massless in the early universe and remains frozen due to Hubble friction. As the temperature drops below 
$T_{\text{osc}}$, the axion (or ALP) field begins to oscillate producing axions and ultimately leading to the 
{chiral gravitational waves.}

\begin{figure}[h]
\includegraphics[scale=0.45]{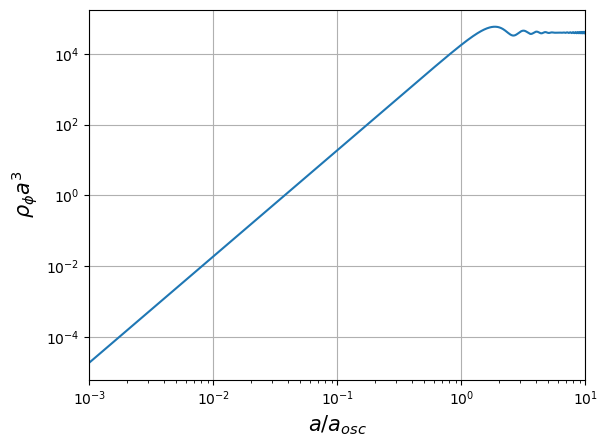}
\caption{\label{fig12}
The axions are produced via the misalignment mechanism, which is a good approximation. We use the model parameters $m=1\times10^{-7}\text{eV}$.}
\end{figure}

During the radiation dominated era, we assume the field $\phi$ to be spatially homogeneous and adopt the comoving Friedmann-Robertson-Walker (FRW) metric $\dd s^{2}=a(\tau)^{2}(-\dd\tau^{2}+\delta_{ij}\dd{x}^{i}\dd{x}^{j})$. With $\phi \ll {\fh} $, the equation of motion (EoM) for the axion field can be expressed as
\begin{equation}\label{axion eom1}
    \phi'' + 2aH \phi' + a^{2}m^{2}\phi = 0.
\end{equation}
The derivative $'$ is with respect to the conformal time $\tau$,  
and the energy density of the axion field is~\cite{Venegas:2021wwm}
\begin{equation}
    \rho_{\phi}=\frac{1}{2}\left(\frac{\dd \phi}{\dd t}\right)^{2}+\frac{1}{2}m^{2}\phi^{2}.
\end{equation}
Given that the comoving axion number $N_{\phi}=n_{\phi}(T)a^{3}$ is conserved, where $n_{\phi}(T)$ is the number density. Furthermore, the energy density of axions can be expressed as $\rho_{\phi}(T)= m(T)n_{\phi}(T)$. Using this relation, the energy density of axions today can be written as follows:
\begin{equation}
    \rho_{\phi}(T) = \rho_{\phi}(T_{\text{osc}})\frac{m(T)}{m(T_{\text{osc}})}\left(\frac{a_{\text{osc}}}{a}\right)^{3}.
\end{equation}
By introducing a misalignment mechanism, we simulate axion generation, with the results of specific calculations shown in Fig.~\ref{fig12}. Although it is not instantaneous when the oscillations begin at $a_{\text{osc}}$~\cite{Venegas:2021wwm}. It is also a good approximation model to facilitate this work. 
\begin{figure}[htbp]
	\centering
	\begin{minipage}{0.49\linewidth}
		\centering
		\includegraphics[width=0.9\linewidth]{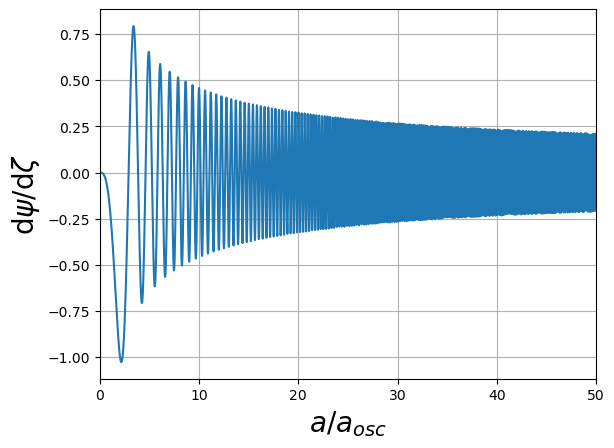}
	\end{minipage}
	\begin{minipage}{0.49\linewidth}
		\centering
		\includegraphics[width=0.9\linewidth]{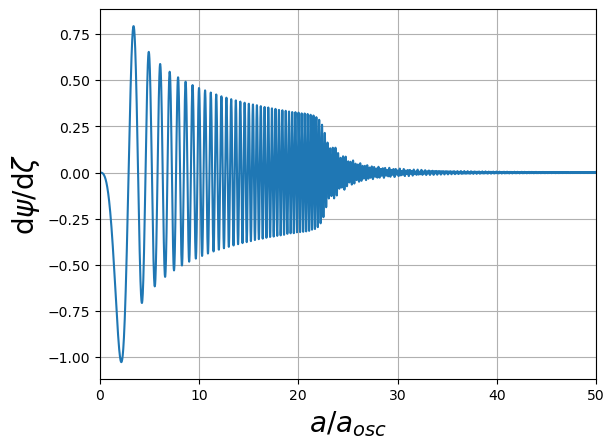}
	\end{minipage}
	\begin{minipage}{0.49\linewidth}
		\centering
		\includegraphics[width=0.9\linewidth]{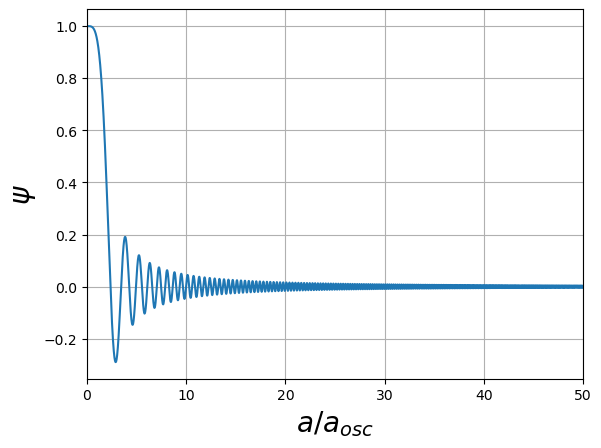}
	\end{minipage}
	\begin{minipage}{0.49\linewidth}
		\centering
		\includegraphics[width=0.9\linewidth]{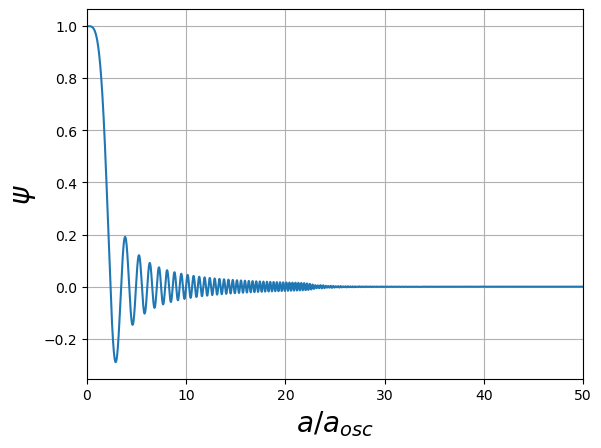}
	\end{minipage}
\caption{\label{backreaction}
The time evolution of normalized field values $\psi$ and velocities of the axion field $\dd\psi/\dd\zz$ with no backreaction in \eqref{axion eom1} (left figures), and with backreaction in \eqref{axion eom} (right figures). We can see that the Nieh-Yan term suppresses the motion of the axion field. The parameters we use are $m=1\times10^{-7}\text{eV}$, $f_{\alpha}=1\times10^{16}\text{GeV}$ and $\alpha=39.08\Mp^{2}$.}
\end{figure}

However, in this study, the Nieh-Yan term affects the evolution of the axion field. By using the action \eqref{total_equation},  the EoM for the axion field in the Nieh-Yan gravity can be obtained as 
\begin{equation}\label{axion eom}
    \phi'' + 2aH \phi' + a^{2}m^{2}\phi = \frac{\alpha a^{2} }{4{\fh}} T_{A\mu\nu} \widetilde{T}^{A\mu\nu}.
\end{equation}
The Nieh-Yan term on the right side of the equation acts as a damping term. In perturbation theory, it can be expressed as a second-order perturbation, and thus the Nieh-Yan term also represents the backreaction of gravitational waves on the background. In Fig. \ref{backreaction}, we demonstrate that when the axion or axion-like particle rolls to its minimum, the oscillations of the axion field are damped by the Nieh-Yan term. This damping implies a significant depletion of the axion’s energy, which is converted into gravitational waves.
Here  $\psi=\phi/\theta {\fh}$ is the dimensionless axion field, {with the initial misalignment angle $\theta=1$}.  The derivative $\dd\psi/\dd\zz$ is defined with  a dimensionless time variable $\zz=ma_{\text{osc}}\tau=\left(a/a_{\text{osc}}\right)$. If the EoM \eqref{axion eom1} lacks a backreaction term, the approximate solution of $\dd\psi/\dd\zz$ can be written as a periodic function. In contrast, the Nieh-Yan term in \eqref{axion eom} will cause the oscillations to be ultimately suppressed. 
The presence of the Nieh-Yan term eventually leads to a significant depletion of axion energy and the production of gravitational waves. Due to the parity-violating mechanism, gravitational waves of different helicities will be produced asymmetrically, which leads to a chiral gravitational wave background. 

\section{Chiral gravitational wave production}
\label{GWP} 
For a more detailed analysis of the mechanism for gravitational wave production, we first consider the cosmological tensor perturbation~\cite{Caprini:2018mtu,Li:2020xjt,Li:2021wij,Izumi:2012qj}. The quadratic action of the tensor perturbation with Nieh-Yan term can be expressed as
\begin{equation}
    S_{(2)} = \int \dd^{4}x \frac{a^{2}\Mp^{2}}{8} \left[h_{ij}'^{2}-\left(\partial_{i}h_{jk}\right)^{2}-\frac{\alpha \phi'}{{\fh} \Mp^{2}}\epsilon^{ijk}h_{il}\partial_{j}h_{kl}\right]. 
\end{equation}
The tensor perturbation $h_{ij}$ can be expanded in the circular polarization basises in Fourier space, expressed as 
\begin{equation}\label{expand}
    h_{ij}(\tau,x) = \sum_{\A =L, R}\int \frac{\dd^3 k}{(2\pi)^{3/2}} e_{ij}^{\A}(\kk) h_{\A}(k, \tau)e^{i \kk\cdot \vec{x}}.
\end{equation}
Here $h_{\A}(k, \tau)$ are the mode functions,  which represent different helicity of the gravitational waves. $e_{ij}^{\A}(\kk)$ is the circular polarisation tensor and $\A =L, R$ denotes the left-handed and right-handed polarizations, respectively.
They satisfy the following relation:
\begin{equation}
\begin{split}
  e_{ij}^{R}(\kk)e_{ij}^{R}(\kk) &= e^{L}_{ij}(\kk)e^{L}_{ij}(\kk) = 0,\\
  e_{ij}^{R}(\kk)e_{ij}^{L}(\kk) &= 2,\\
  ik^{p}\epsilon_{mpj} e^\A_{ij} (\kk) &= \lambda_{\A} k \ e^{\A}_{mi}(\kk) \ (\A=L, R). 
\end{split} \label{cptr}
\end{equation}
The value of  $\lambda_{\A}$ can be either $+1$ or $-1$, corresponding to the left-handed and right-handed modes, respectively. By using  these relations, the EoM of the mode functions $h_{\A}(k,\tau)$ can be obtained as follows:  
\begin{equation}\label{mode function}
\begin{split} 
&h_{\A}''(k,\tau) + 2aHh_{\A}'(k,\tau) + 
\omega_{\A}^{2} h_{\A}(k,\tau) =0,\\
&\omega_{\A}^{2}\equiv k^{2} \pm\frac{\alpha \phi'(\tau)}{{\fh}\Mp^{2}}k,\quad\A =L, R.
    \end{split} 
\end{equation}
Through analytical analysis of the mode function, we can see that the solution depends on the time-dependent function $\omega^{2}_{\A}$. As the field $\phi(\tau)$  begins rolling, the sign of $\phi'(\tau)$ undergoes persistent alternations. Consequently, one helicity mode corresponds to a negative $\omega_{\A}^{2}$, which means the tachyonic instability. This instability drives an exponential growth of gravitational wave modes $h_{\A}(k,\tau)\sim e^{\left|\omega_{\A}\right| \tau}$.
Since the growth rate is related to helicities, the left-handed and right-handed gravitational waves are generated asymmetrically, ultimately leading to chiral gravitational waves.


Then we consider the cosmological tensor perturbation of the dumping term in the right-hand side of the EoM~\eqref{axion eom}. The Nieh-Yan term resembles the term associated with the Kalb-Ramond field~\cite{Chatzistavrakidis:2021oyp,Sorkhi:2023bfj}. By the linearized order, the EoM~\eqref{axion eom} becomes
\begin{equation}\label{axion eom modify1}
\begin{split}
{\phi''} + 2aH \phi' + a^{2}m^{2}\phi = 
    \frac{\alpha a^{2}}{4{\fh}}( \epsilon^{\mu\nu\rho\sigma}\partial_{\mu}h_{\nu\lambda}\partial_{\rho}h^{\lambda}_{~\sigma}\\
+2\epsilon^{\mu\nu\rho\sigma}\partial_{\mu}b_{\nu\lambda}\partial_{\rho}h^{\lambda}_{~\sigma}+\epsilon^{\mu\nu\rho\sigma}\partial_{\mu}b_{\nu\lambda}\partial_{\rho}b^{\lambda}_{~\sigma}). 
\end{split}
\end{equation}
We are interested in the {Nieh-Yan term modified TEGR}, where the Kalb-Ramond perturbation $b_{\mu\nu}=0$. By decomposing the remaining terms and retaining only the non-zero components, the EoM of the axion field can be simplified to
\begin{equation}\label{axion eom modify2}
    \phi'' + 2aH \phi' + a^{2}m^{2}\phi = 
    \frac{\alpha a^{2}}{4{\fh}}  \epsilon^{ijk}\partial_{0}h_{il}\partial_{j}h_{kl}.
\end{equation}
The term on the right side of the equation means that the gravitational waves will induce a geometric backreaction on the axion field. Combining it with equation~\eqref{mode function}, we can see that the nature of the damping term is the gravitational wave spatial perturbations, which means a new effective mechanism to reduce axion energy density. It is remarkable that, in a given parameter region, this mechanism could reduce the energy density of axions to today's dark matter relic abundance. These results will be presented in the following sections.

\subsection{Numerical calculation}
\label{numerically calculated}
In this subsection, we discuss how to obtain our results through numerical calculation. During the radiation-dominated era, we have $ \frac{a}{a_{\text{osc}}}=\frac{\tau}{\tau_{\text{osc}}}$.
The tachyonic instability can occur when the comoving time exceeds the oscillation time $\tau_{\text{osc}}=1/ma_{\text{osc}}$, with ${H}_{\text{osc}}\sim m$. This allows us to introduce a new time variable:
\begin{equation}\label{zzz}
    \zz \equiv \frac{a}{a_{\text{osc}}} =\frac{\tau}{\tau_{\text{osc}}}= ma_{\text{osc}}\tau.
\end{equation}
Then we introduce the auxiliary field $v_{\A}(k,\tau) \equiv {\Mp}a(\tau)h_{\A}(k,\tau)$, from \eqref{zzz} we have  $a''=0$, the EoM \eqref{mode function} of mode function for the gravitational waves becomes~\cite{Kachelriess:2022jmh}
\begin{equation}\label{vdd}
    v_{\A}''(k,\tau) + \left(k^{2} \pm\frac{\alpha \phi'}{{\fh}\Mp^{2}}k\right)v_{\A}(k,\tau) = 0.
\end{equation} 

During the radiation-dominated era, the approximate solution of the scalar fields in Eq.~\eqref{axion eom modify2} is $\phi(t)\simeq \phi_{\rm osc}\zz^{-3/2}\cos(mt)$, 
and $|\phi'|\simeq \theta \fh\zz^{-3/2} m a$ \cite{Machado:2018nqk}.
It is worth noting that the tachyonic instability in Eq.~\eqref{vdd} depends on the sign of  $\phi'$, which changes every half period.
Thus, the occurrence of exponential growth requires the growth timescale $\left|\omega_\mathcal{A}\right|^{-1}$
is less than the conformal oscillation time $\left(m a\right)^{-1}$.
By solving $ k^{2}_\pm \pm\frac{\alpha \phi'}{{ f_\alpha} M_{Pl}^{2}}k_\pm=-(ma)^{2}$, we can obtain the tachyonic production band as $k_{-}<k<k_{+}$,  where $k_{\pm}$ are
\begin{equation}\label{band}
    k_{\pm} \approx \widetilde{k}\left[1\pm \sqrt{1- \zz^{3}\left(\frac{2\Mp^{2}}{\alpha\theta}\right)^{2}}\right],
\end{equation}
and $\widetilde{k}=\frac{\alpha\theta}{2\Mp^{2}}\frac{ma_{\text{osc}}}{\zz^{1/2}}$. This shows that $|\alpha\theta|>2\Mp^{2}$ is necessary for the tachyonic production band to be initially open, and the band will close at the time when $\zz=\zz_\alpha\equiv\left(\alpha\theta/2\Mp^{2}\right)^{2/3}$.
However, equation \eqref{band} neglects the suppression of the motion of the axion field by the Nieh-Yan term, causing the approximate band close time $\zz_\alpha$ earlier than the actual value. In fact, the band transitions from a broadband to a narrow band, in which narrow resonances affect energy density at high frequencies. 

By defining the dimensionless momentum $\hat{k}=k/ma_{\text{osc}}$,  equation~\eqref{vdd} can be expressed as a Mathieu equation~\cite{Kofman:1997yn,Dufaux:2006ee,mclachlan1947theory} 
\begin{equation}\label{phi2}
  \frac{\dd^{2}v_{\A}}{\dd\zz^{2}}(k,\tau) + \left(\hat{k}^{2} \pm \frac{\alpha\hat{k}}{{\fh}\Mp^{2}}\frac{\dd\phi}{\dd\zz}\right)v_{\A}(k,\tau)=0. 
\end{equation}

Introducing another auxiliary dimensionless field $\eta_{\A}(k,\tau)=\sqrt{ma_{\text{osc}}}v_{\A}(k,\tau)$ and dimensionless scalar field $\psi=\frac{\phi}{\theta f_{\alpha}}$, the EoM for the mode functions of gravitational waves can be expressed as
\begin{equation}\label{nc_grivton}
    \frac{\dd^{2}\eta_{\A}}{\dd\zz^{2}}(k,\tau)+\left(\hat{k}^{2}\pm \frac{\alpha\hat{k}\theta}{\Mp^{2}}\frac{\dd\psi}{\dd\zz}\right)\eta_{\A}(k,\tau)=0.
\end{equation}

Next, using \eqref{zzz}, we can transform the EoM \eqref{axion eom modify2} into the new time variable $\zz$, resulting in 
\begin{equation}\label{simplify eom}
    \frac{\dd^{2}\phi}{\dd\zz^{2}} + \frac{2}{\zz}\frac{\dd\phi}{\dd\zz} +\zz^{2}\phi = \frac{\alpha}{4{\fh}} \frac{\zz^{2}}{m^{2}}\epsilon^{ijk}\partial_{0}h_{il}\partial_{j}h_{kl} . 
\end{equation}
Assuming that the axion field is homogeneous, we treat the axion field as a mean field and expand the geometric backreaction term using \eqref{expand}. 
By employing the dimensionless wave number $\hat{k}$, the dimensionless time variable $\zz$, and dimensionless scalar field $\psi$ and calculating the integral in the spherical coordinate system, the equation \eqref{simplify eom} can be rewritten as
\begin{equation}\label{nc_axion}
\begin{split}
&\frac{\dd^{2}\psi}{\dd\zz^{2}} + \frac{2}{\zz}\frac{\dd\psi}{\dd\zz} + \zz^{2}\psi \\
=&\frac{\alpha}{8\pi^{2}\fh^{2}\theta} \frac{m^{2}}{\zz^{2}\Mp^2}\sum_{\A=R,L} \int \dd\hat{k} \hat{k}^{3} \Big[\frac{\partial \left|\eta_{\A}\right|^{2}}{\partial \zz}\Big]. 
\end{split}
\end{equation}

For the initial condition, we choose the  Bunch-Davies vacuum.
The mode functions of the gravitational waves can be expressed as
$v_{\A}(\tau) = \frac{1}{\sqrt{2k}}e^{-ik\tau}$, which implies
$\eta_{\A}(\tau) = \frac{1}{\sqrt{2\hat{k}}}e^{-i\hat{k}\zz}$.
The initial amplitude of the oscillation of the axion field is determined by the misalignment angle $\theta$, such that $\phi_{\text{osc}}=\theta {\fh}$ with $\theta\sim O(1)$, which implies $\psi_{\text{osc}}=\frac{\phi_{\text{osc}}}{\theta f_{\alpha}}\sim 1$. We assume that the initial velocity $\phi'$
is negligible, and the scalar field energy density is sufficiently small compared with the radiation bath, which allowed us to ignore its backreaction on the FRW spacetime during the radiation-dominated era.

Finally, we employ the finite difference method to divide $\hat{k}$ space into more than $10^{3}$ segments, $\zz$ into over $10^{4}$ segments, using more stable and precise algorithms to ensure numerical stability and to reveal more details of the mechanism.

\subsection{Stochastic GW background}\label{SBGW}
\label{s4}
We have seen that the Nieh-Yan term causes the chiral gravitational waves. These gravitational waves will all eventually evolve into the stochastic gravitational wave background (SGWB) and manifest in our current universe. For SGWB, the energy spectral density of it can be expressed as~\cite{Caprini:2018mtu}
\begin{equation}\label{omega_GW}
\Omega_{\GW} = \frac{1}{\rho_{\text{c}}}\frac{\dd\rho_{\GW}}{\dd \log k}.
\end{equation}
Here, $\rho_{\text{c}}$ is the critical density during the radiation-dominated era, which is defined as
\begin{equation}\label{rho_c}
    \rho_{\text{c}} = 3\Mp^{2}H^{2} = 3\Mp^{2}m^{2}\left(\frac{a_{\text{osc}}}{a}\right)^{4}. 
\end{equation}

\begin{figure}[h]
\flushleft
\includegraphics[scale=0.5]{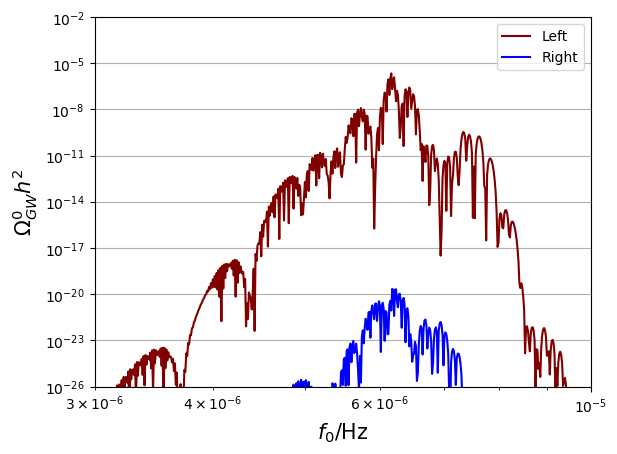}
\caption{\label{fig1}
 The energy spectral density of the gravitational wave for ALP2 with parameters in Table \ref{table1}. The red and blue lines correspond to the spectrum for left-handed and right-handed gravitational waves.}
\end{figure}

Furthermore, we can obtain the specific form of  gravitational wave energy density  in  energy-momentum tensor, which can be written as 
\begin{equation}\label{ED1}
  \frac{\dd \rho_{{\GW}}}{\dd \log k}
  =\frac{\Mp^2k^3}{8\pi^2a^2}\mathcal{P}_{h'}(k, \tau)=\frac{\Mp^2 k^3}{4 \pi^2 a^2}\sum_{\A=R, L}\left|h_{\A}'\right|^{2}.     
\end{equation}
Then, consider dimensionless momentum $\hat{k}$, auxiliary field $\eta_{\A}$, the energy spectral density of gravitational waves in \eqref{omega_GW} can be rewritten as
\begin{equation}\label{ED3}
    \Omega_{\GW} = \frac{\hat{k}^{3}m^{2}}{6\pi^{2}\Mp^{2}}\sum_{\A=R, L}\left|\frac{\dd\eta_{\A}}{\dd\zz}\right|^{2}.  
\end{equation}

By numerically solving EoMs \eqref{nc_grivton} and \eqref{nc_axion}, and substituting the results into \eqref{ED3}, we can obtain the total energy spectral density of chiral gravitational waves. Finally, we need to obtain the amplitude and frequency of the gravitational wave spectrum today, which can be expressed as
\begin{equation}
\begin{split}
    \Omega^{0}_{\GW} &= \Omega^{*}_{\GW} \left(\frac{g_{\rho,*}}{g^{\gamma}_{\rho,0}}\right)\left(\frac{g_{s,\text{eq}}}{g_{s,*}}\right)^{\frac{4}{3}}\Omega^{0}_{\gamma} \\
    &\approx  1.67 \times 10^{-4} g^{-1/3}_{\rho,*} \Omega_{\GW}^{*},
\end{split}
\end{equation}
where $g_{s,\text{eq}}=2+2N_{\text{eff}}\left(\frac{7}{8}\right)\left(\frac{4}{11}\right)=3.938$, $\Omega_{\gamma}^{0}=5.38\times10^{-5}$, and $\Omega_{\GW}^{*}$ is the energy spectral density {when gravitational waves stopped growing.}
Moreover, $g_{\rho}$ is the number of effective degrees of freedom associated with the energy density. We assume $g_{\rho,0}^{\gamma}=2$ and use the approximation that the effective degrees of freedom at the moment of gravitational wave generation $g_{\rho,*}=g_{s,*}$. 
Because the mechanism occurs approach the QCD phase transition, we choose $g_{s,*}=106.75$.

Taking into account redshift, the frequency of gravitational waves in the present era can be written as
\begin{equation}\label{f_0}
    f_{0} = \frac{k}{2\pi a_{0}} \simeq 7.125 \times 10^{-4} \text{Hz}\left(\frac{100}{g_{\rho,*}}\right)^{\frac{1}{12}} \left(\frac{k}{ma_{\text{osc}}}\right) \left(\frac{m}{\text{eV}}\right)^{\frac{1}{2}}.
\end{equation}
The final numerical calculations with the parameters of ALP2 are presented in Fig.~\ref{fig1}. It is intuitively clear that, since the growth rate is related to helicities, there is a difference in the magnitude of the energy density between the left-handed and right-handed gravitational waves.

\begin{table}[h]
\centering
\begin{tabular}{|c|c|c|c|c|c|}
\hline 
Models      &   m(eV)     &${\fh}$(GeV)   &$\alpha$($\Mp^{2}$) &$\rho^{0}_{\phi}/\rho^{0}_{DM}$  &$\Delta N_{\text{eff}}$  \\
\hline 
QCD1   &$5.69\times10^{-11}$    &$1\times10^{17}$    &42.63   &8.9   &0.138 \\
\hline 
QCD2   &$2.85\times10^{-10}$    &$2\times10^{16}$    &43.16   &0.7   &0.003 \\
\hline 
ALP1   &$1\times10^{-7}$        &$1\times10^{16}$    &40.11   &1.6   &0.0014\\
\hline 
ALP2   &$1\times10^{-6}$        &$1\times10^{16}$    &39.08   &3.9   &0.0012\\
\hline 
ALP3   &$1\times10^{-3}$        &$1\times10^{16}$    &36.53   &91.3    &0.0009\\
\hline 
ALP4   &$1\times10^{-1}$        &$1\times10^{17}$    &35.61   &-   &0.059\\
\hline 
ALP5   &$1\times10^{0}~$         &$1\times10^{17}$    &34.85   &-   &0.067\\  
\hline
\end{tabular}
\caption{{Parameters of axion and axion-like particles, we have chosen $\theta=1$ in all the numerical calculations.}\label{table1}}
\end{table}

The chirality of chiral gravitational waves can be described by the degree of circular polarization, which can be expressed as 
\begin{equation}
    \Pi = \frac{\Omega_{\text{GW}}^R-\Omega_{\text{GW}}^{L}}{\Omega_{\text{GW}}^{R}+\Omega_{\text{GW}}^{L}}.   
\end{equation}
As shown in Fig.~\ref{fig1}, chiral gravitational waves exhibit significant polarization in the peak interval, indicating that purely left-handed gravitational waves are the dominant contribution to the total energy density.

\section{Parameter and  detectability analysis}\label{detection}

In this section, we explore the parameter space of axion masses and decay constants for detectable chiral gravitational wave signals, then fit them with the single peak curve.  
The parameters we used are displayed in Table~\ref{table1}.

\begin{figure}[h]
\includegraphics[scale=0.5]{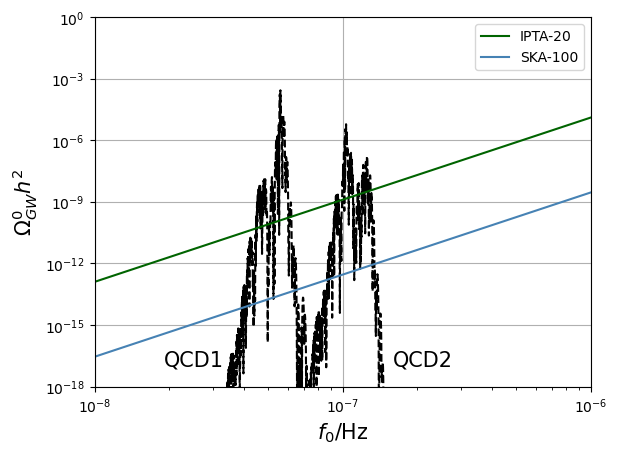}
\caption{\label{fig3}
The energy spectral density of nanohertz chiral gravitational wave background from QCD1 and QCD2 with parameters in Table \ref{table1}, as well as the sensitivity curves of IPTA-20~\cite{Antoniadis:2022pcn,Shannon:2015ect}(green line) and SKA-100~\cite{Carilli:2004nx,Kuroda:2015owv}(blue line).}
\end{figure}
In Fig.~\ref{fig3}, the nanohertz chiral gravitational waves are explored and compared with the PTA observations, including SKA and IPTA. The gravitational wave signals correspond to two parameters in the QCD axion region. Two black curves represent the energy spectral density of QCD1 and  QCD2 with the parameters in Table \ref{table1}. We can see that the relic density of QCD1(QCD2) is higher(lower) than the observed dark matter relic energy density.

\begin{figure}[h]
\includegraphics[scale=0.5]{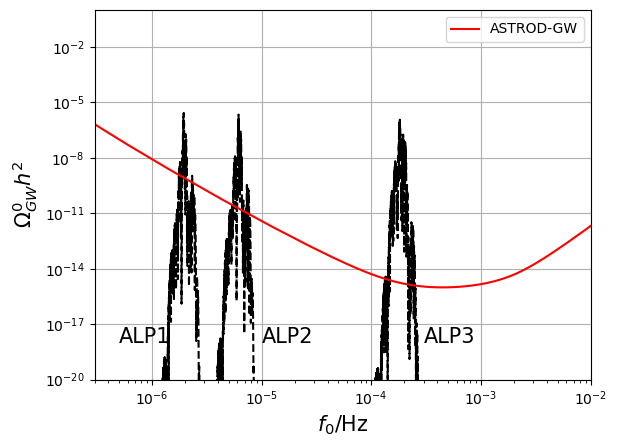}
\caption{\label{fig5}
The energy spectral density of microhertz chiral gravitational wave energy background from ALP1, ALP2, and ALP3 with the parameters in Table~\ref{table1}, as well as the sensitivity curve of ASTROD-GW~\cite{Kuroda:2015owv}(red line).
}
\end{figure} 

In Fig.~\ref{fig5}, the microhertz chiral gravitational waves are explored and compared with the ASTROD-GW~\cite{Kuroda:2017znl}.  Three black curves represent the energy spectral density of ALP1, ALP2, and ALP3 with the parameters in Table \ref{table1}. We can see that the relic density of these ALPs is higher than the observed dark matter relic energy density.

\begin{figure}[h]
\includegraphics[scale=0.5]{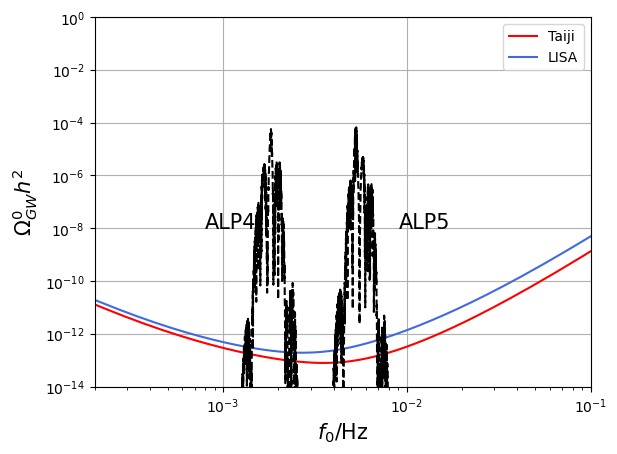}
\caption{\label{fig4}
The millihertz chiral gravitational wave energy spectral density of ALP4 and ALP5 with parameters in Table~\ref{table1}, as well as the sensitivity curves of LISA~\cite{Caprini:2019pxz}(blue) and Taiji~\cite{Liu:2023qap}(red).  
}
\end{figure}

In Fig.~\ref{fig4}, the millihertz chiral gravitational waves are explored and compared with the LISA and Taiji.  Two black curves represent the energy spectral density of ALP4 and ALP5 with the parameters in Table \ref{table1}. We can see that the relic density of ALP4(ALP5) is too high than the observed dark matter relic energy, however, other mechanisms may need to be
considered ~\cite{Machado:2018nqk,Madge:2021abk}.

\begin{figure}[ht]
\includegraphics[scale=0.5]{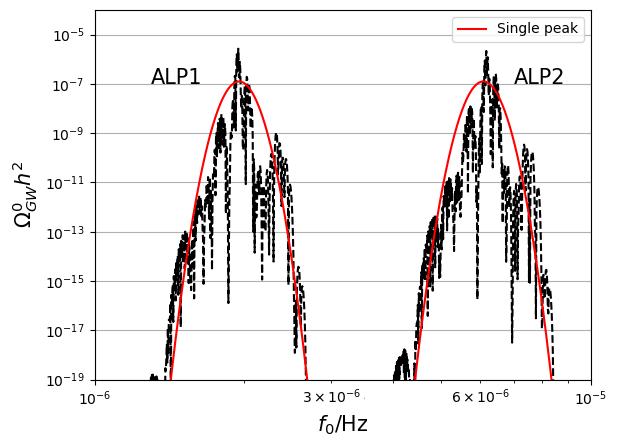}
\caption{\label{fig15}
The gravitational wave energy spectral density (black curves) of ALP1 (left) and ALP2 (right) with parameters in Table~\ref{table1}, which are fitted by the red line with the single peak curve in~\eqref{single peak}.}
\end{figure}

To facilitate model-independent tests,  we use the single peak curve to fit the energy spectral density of the gravitational waves. The single peak curve can be expressed as follows~\cite{Kohri:2018awv,Inomata:2020tkl,Cai:2019amo,White:2021hwi,Lozanov:2022yoy}:
\begin{equation}\label{single peak}
    \Omega_{\text{sp}}(f_0) = \Omega_{c}\exp\left[- \frac{\log^2 (f_0/f_{\text{c}})}{\sigma^2}\right].
\end{equation}
The characteristic value of the energy density $\Omega_{c}=c_{\text{eff}}\Omega_{\phi}$, where 
$c_{\text{eff}}$ is the characteristic conversion efficiency, and
\begin{equation}
    \Omega_{\phi} = \frac{\theta^{2}\fh^{2}/2}{3\Mp^{2}} \frac{m^{2}}{H_{\text{osc}}^{2}}\sim \left(\frac{\theta {\fh}}{\Mp}\right)^{2}.
\end{equation}
{We can get the characteristic frequency by approximate analysis Eqs. \eqref{band} and \eqref{f_0}} 
\begin{equation}   
f_{c}=7.125 \times 10^{-4} \text{Hz}\left(\frac{100}{g_{*}}\right)^{\frac{1}{12}} \left(\frac{9}{14}\frac{\alpha\theta}{\Mp^{2}}\right)^{\frac{2}{3}}\left(\frac{m}{\text{eV}}\right)^{\frac{1}{2}}.
\end{equation}
In Fig.~\ref{fig15}, the gravitational wave energy spectral densities of ALP1 and ALP2 are fitted with the single peak curve. Using semi-analytical fitting, the parameters in \eqref{single peak} can be obtained as $\sigma\simeq 0.06$ and $c_{\text{eff}} \simeq 2 \times 10^{-2}$.

\begin{figure}[h]
\includegraphics[scale=0.5]{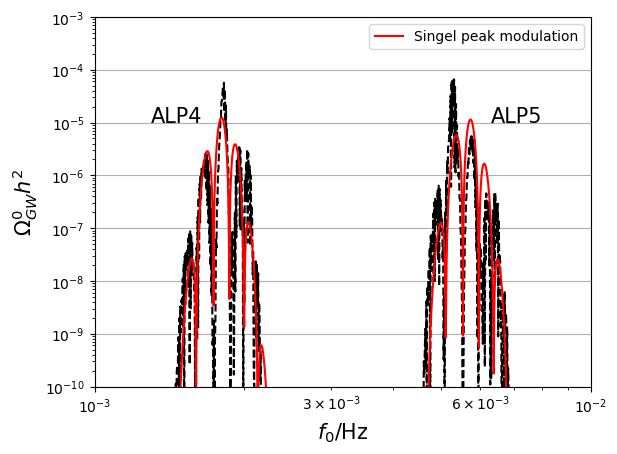}
\caption{\label{fig16}
The gravitational wave energy spectral density (black curves) of ALP4 (left) and ALP5 (right) with parameters in Table~\ref{table1}. They are fitted by the red line with single peak curve modulation in~\eqref{modified single peak curve}.}
\end{figure}

To achieve a more accurate fit, the single peak curve~\eqref{single peak} can be improved with modulation as 
\begin{equation}\label{modified single peak curve}
    \Omega_{\text{spm}}(f_0) =
    \Omega_{\text{sp}} (f_0) \text{cos}^{2}\left(\beta\frac{f_0}{f_{\text{c}}}+\xi \right).
\end{equation}
In Fig.~\ref{fig16}, the gravitational wave energy spectral densities of ALP4 and ALP5 are fitted by the single peak curve modulation  \eqref{modified single peak curve}  with the red lines. We find that $\beta=1.2\alpha\theta/\Mp^{2}$, and a constant phase $\xi=-1.8$ provides a good approximation.

The axion-like mechanism with gravitational coupling through the Nieh-Yan term can effectively reduce the axion energy and generate a chiral gravitational wave background. The gravitational waves will alter the number of effective relativistic degrees of freedom $N_{\text{eff}}$, and the effective contribution is given by
\begin{equation}\label{eff}
    \Delta N_{\text{eff}} = \frac{8}{7}\left(\frac{11}{4}\right)^{\frac{4}{3}}\frac{\rho_{\text{GW}}}{\rho_{\gamma}},
\end{equation}
where $\rho_{\gamma}$ is the photon energy density. Ignoring the temperature dependence of the mass, $\Delta N_{\text{eff}}$ can be approximately expressed as~\cite{Agrawal:2017eqm}
\begin{equation}
    \Delta N_{\text{eff}} \sim \frac{g_{\rho,*}\pi^{2}}{90}\frac{\theta^{2}\fh^{2}}{\Mp^{2}}\frac{a_{\text{b}}}{a_{\text{osc}}},
\end{equation}
where $a_{\text{b}}$ is the scale factor at which gravitational wave backreaction becomes significant for the axion.  The effective relativistic degrees of freedom have been constrained at $95\%$ confidence level to $\Delta N_{\text{eff}} \textless 0.3$~\cite{Planck:2018vyg}. 
In Table \ref{table1}, we numerically calculate $\Delta N_{\text{eff}}$ using equation \eqref{eff}, the specific values displayed remain below this upper limit.

\begin{figure}[htbp]
\includegraphics[scale=0.5]{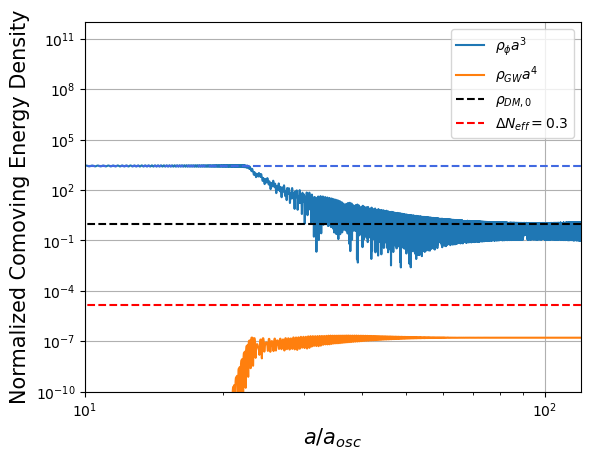}
\caption{{The evolution of normalized comoving energy densities of axions $\rho_{\phi}a^{3}/\rho_{\text{DM},0}$(blue line) and gravitational waves $\rho_{\text{GW}}a^{4}/\rho_{\text{DM},0}$(orange line) 
with the parameters of QCD2 in Table~\ref{table1}. The royal blue dashed line represents the $\rho_{\phi}a^{3}/\rho_{\text{DM},0}$ with non-backreaction, and the black dashed line represents today's dark matter energy density $\rho_{\text{DM},0}/\rho_{\text{DM},0}$. The Red dashed line means the upper bound which is restricted by $\Delta N_{\text{eff}}=0.3$.} 
\label{fig2}}
\end{figure}

In Fig.~\ref{fig2}, we illustrate the evolution of normalized comoving energy densities with the optimal parameters of QCD2, which reduce the axion comoving energy density lower than current dark matter relic density and satisfy the constraints on effective relativistic degrees of freedom. 
It is evident that during the initial stage, the backreaction is weak, the oscillations remain stable, and the suppression of the comoving energy density of the axion is not significant. As the gravitational wave energy density increases, the backreaction effect gradually dominates, leading to the suppression of the comoving energy density of axions.

\section{Conclusion and Discussion}
\label{conclusion}

In this work, we use the Nieh-Yan term to couple with the axion or axion-like particle field, resulting in the generation of a chiral gravitational wave background. We discuss how to build the model and how the axion dynamics are affected by the geometric backreaction of the Nieh-Yan term. Then, we numerically calculate the chiral gravitational wave energy spectral density. We explore the parameter space defined by the axion mass and decay constants, which could be detected by PTA or future space-based gravitational wave detectors. Specifically, we compare energy spectral density with the sensitivity curves of current and future gravitational wave detection projects.

Furthermore, we numerically calculate the evolution of axion and gravitational wave comoving energy density over time. Our findings suggest that this mechanism offers a new, efficient way to consume the energy density of axion dark matter. Under certain parameters, the eventual relic density of axions can be comparable to the relic density of dark matter today. Also, within the explored parameter space, the contribution to the effective degrees of freedom from the eventual production of gravitational radiation remains within observational limits.

It has been shown that the PTA is not sensitive to the chirality of SGWB signals~\cite{Kato:2015bye}. However, with the supplement of astrometry, it has the potential to detect the chirality of SGWB in the nanohertz band~\cite{Liang:2023pbj,Caliskan:2023cqm}. Because the gravitational wave signals affect the precise measurement of stellar positions. By measuring the non-vanishing EB correlation within the two-point correlation function of stellar positional excursions, it is possible to detect the chirality. Such measurements, in turn, complement PTA's detection of the SGWB with chirality.

On the other hand, in space-based gravitational wave detections, a single detector is not sensitive to the chirality of SGWB signals. This is due to the mirror symmetry introduced by the planar structure of a single space interferometer, which causes the chirality cancelation of gravitational wave signals with uniform and isotropic backgrounds. Fortunately, with the advancement of space-based gravitational wave detection projects, such as LISA, Taiji~\cite{Hu:2017mde} and TianQin~\cite{TianQin:2015yph}, it is now possible to detect the chirality and polarizations through a joint network of the detectors, which helps avoid chiral cancellation that may occur with a single detector~\cite{Seto:2020zxw,Liu:2022umx, Chen:2024xzw, Chen:2024ikn,Chen:2024fto,Su:2025nkl}.  The study on the response of this chiral signal from audible axion based on a network of space-based gravitational wave detectors is in preparation.

\begin{acknowledgments}
This work is supported by 
the National Key Research and Development Program of China (Nos. 2023YFC2206200, 2021YFC2201901), the National Natural Science Foundation of China (No.12375059) and the Fundamental Research Funds for the Central Universities (No. E2ET0209X2).
We thank Da Huang and Gang Wang for many valuable discussions.
\end{acknowledgments}

\bibliography{myrefs.bib}

\end{document}